\documentclass[12pt]{iopart}

\usepackage{graphicx}
\usepackage{tabularx}
\usepackage{color}
\usepackage[normalem]{ulem}

\begin{document}

\title[]{Vacancy and Antisite-Induced Ferromagnetism in Liquid-Phase Exfoliated Bi$_2$Te$_3$}

\author{V. Gomez$^1$, M.J. Saenz$^1$, J. Canaval$^1$, L. Cuadrado$^1$, C. Espejo$^2$, W. Lopez$^2$, Y. Hernandez$^1$}

\address{$^1$Nanomaterials Lab, Physics Department, Universidad de los Andes, 111711, Bogota, Colombia}
\address{$^2$Department of Physics and Geosciences, Universidad del Norte, Barranquilla, Colombia}
\ead{yr.hernandez@uniandes.edu.co}
\vspace{10pt}
\begin{indented}
\item[]September 2024
\end{indented}

\begin{abstract}

Bismuth Telluride (Bi$_2$Te$_3$) is a widely studied topological insulator, recognized for its unique surface states, low electronic bandgap, and low thermal conductivity. In this study, we characterize exfoliated Bi$_2$Te$_3$ dispersions produced via solvothermal intercalation, where ferromagnetism was measured at room temperature. DFT simulations show that this ferromagnetic behavior is attributed to the presence of vacancies and antisites in both the bulk material and the exfoliated crystals. Additionally, the DFT results were complemented by experimental measurements of the optical bandgap using UV-Vis spectroscopy, revealing a broadening of the bandgap as the material becomes thinner.


\end{abstract}

%
%
%
%
%

\section{Introduction}

With the discovery of graphene and the development of liquid-phase exfoliation methods for layered materials, the nanostructuring of bismuth telluride (Bi$_2$Te$_3$), a topological insulator known for its promising thermoelectric properties, has been revisited as an economical and scalable method to obtain nanoscale crystals. The introduction of vacancies and antisites in this material has demonstrated potential for tuning its magnetic and optical characteristics, making it an ideal candidate for future technological advancements, such as energy-efficient spintronics devices.\\

Xiao et al. measured for the first time a ferromagnetic signal at room temperature in nanostructured hierarchical architectures (HAs) of Bi$_2$Te$_3$ \cite{xiao2014unexpected}. These caterpillar-like HAs, with dimensions up to the micrometer scale, were prepared using airless hydrothermal synthesis and without the introduction of exotic magnetic dopants, which were typically used in this material to induce such magnetic properties \cite{niu2011mn, lee2014ferromagnetism}. Their results revealed that intrinsic point defects, specifically antisite Te sites, were responsible for the creation of a magnetic moment, providing new insights into the origins of magnetism in topological insulators. In this work, the measurement of a ferromagnetic signal is reported to be within the same order of magnitude as previously observed, but in Bi$_2$Te$_3$ samples produced from commercially available powder with distinct structural characteristics. Specifically, the current study focuses on low-dimensional crystals containing fewer than three quintuple layers.\\

In recent years there has been considerable interest in studying the thermoelectric and topological properties of Bi$_2$Te$_3$ under several conditions using first-principles calculations. Reports of the electronic structure and topological properties of the Bi$_2$Te$_3$ crystal are well known, where the location of band edges and the effective mass parameters for electrons and holes associated with these band edges are obtained \cite{wang2007electronic,luo_first-principles_2012,lawal2019electronic}. Other authors have calculated the band dispersion and spin texture of topologically protected surface states in the bulk topological insulators Bi$_2$Se$_3$ and Bi$_2$Te$_3$ and their thermoelectric properties by first-principles methods [4,5]. Phonons of single quintuple layers of Bi$_2$Te$_3$ and Bi$_2$Se$_3$ and the corresponding bulk materials were calculated by ab-initio methods from which some features in the Raman measurements of these materials have been explained [6]. The above mentioned work from Xiao et. al. also presents the earliest results from computational studies of magnetism in Bi$_2$Te$_3$ induced by native point defects. It was found that the most prominent native point defects are the antisite defects on the Te$_{(1)}$ and Te$_{(2)}$ sublattices of the Bi$_2$Te$_3$ structure. In fact, these calculations allowed for the identification of the Bi-Te$_{(2)}$ antisite as the trigger of the magnetization in this system \cite{xiao2014unexpected}. Other studies present theoretical results of the thermoelectric and magnetic behavior of Bi$_2$Te$_3$ with vacancy defects and antisites in the Te$_{(1)}$ and Te$_{(2)}$ sublattices, with results comparable to experimental data \cite{tang2019bi}. \\

In the present study, both experimental and computational methods are combined to investigate the effects of vacancies and antisite defects in liquid-phase exfoliated Bi$_2$Te$_3$, providing new insights into its magnetic and optical behavior. The most significant finding was the emergence of a ferromagnetic signal at room temperature, as well as the broadening of the optical bandgap in the liquid-phase exfoliated material. An analysis of the optical behavior of Bi$_2$Te$_3$ after exfoliation was conducted, opening a previously unexplored discussion on the magnetic and electronic properties of a polydisperse mixture. Our results pave the way for engineering topological insulators with ferromagnetic properties at room temperature, which represent a promising advancement in current materials physics \cite{chang2015high, katmis2016high}. \\

\section{Methods}

The most common methods for the exfoliation of 2D materials are mechanical cleavage and liquid-phase processing. The choice of method depends on the desired properties of the material and available resources \cite{6}. Liquid-phase exfoliation (LPE) is a well-established and versatile technique for producing low-dimensional nanomaterials by separating the bulk materials layers into individual sheets using solvents or surfactants \cite{7, hernandez2008high}. In this method, the bulk material is first dispersed in a suitable liquid medium, followed by ultrasonication to aid the exfoliation process. The resultant dispersion can then be processed to remove thicker crystals and control the size and thickness of the nanosheets \cite{7, hernandez2008high}. The method can be used for a wide range of materials, including transition metal dichalcogenides, layered metal oxides and graphene.This technique has been shown to be effective in producing high-quality, stable dispersions of Bi$_2$Te$_3$ nanosheets with a high aspect ratio and a thickness of a few nanometers \cite{9}. Moreover, the LPE technique allows for the control of the thickness and lateral size of the nanosheets by adjusting the ultrasonication time and power. The choice of solvent plays a crucial role in the quality of the resulting nanosheets. For example, polar solvents such as N-methyl-2-pyrrolidone (NMP) and dimethylformamide (DMF) are commonly used to obtain high-quality Bi$_2$Te$_3$ nanosheets with a high aspect ratio \cite{5,10}.

\subsection{Bi$_2$Te$_3$ exfoliation and sample preparation}
A solvothermal intercalation/exfoliation process based, with minor modifications, on the process reported by Ren. L, et al.\cite{ren_large-scale_2012} was implemented. A typical solvothermal reaction was carried out within a Teflon lined vessel mixing 50 mg of lithium hydroxide (LiOH) and 50 mg of Bi$_2$Te$_3$ powder (Sigma-Aldrich, 325 mesh, 99.99\%) in ethylene glycol (15 mL). The autoclave was then heated for 24 h at 200°C. After cooling the resulting solid was purified by membrane filtering (PVDF pore size 0.1mm) and subsequently washed with purified water (MilliQ) and isopropanol (190764 Merck) to remove the excess of lithium hydroxide and water, respectively. Once filtered, the expanded powder was dispersed via bath sonication (Branson 1800, 6 hours at room temperature) in 15 mL of a 70:30 N-Methyl-2-pyrrolidone (NMP) and Milli-Q water mixture, and left to decant for 24 hours to subsequently recover the supernatant. The obtained dispersion was vacuum filtered using a PVDF filter with a pore size of 0.1 mm. Then, the material deposited on the filter was carefully washed with acetone and dried to remove any solvent. Finally, the obtained powder was scraped off the filter and prepared for structural and magnetic measurements.

\subsection{Computational modeling}

Pristine bismuth telluride (Bi$_2$Te$_3$) crystallizes in a rhombohedral structure with five atoms per unit cell within the $D^5_{3d}$ ($R\bar{3}m$) space group. The system can also be visualized as a hexagonal unit cell with a layered structure along the z-axis, containing fifteen atoms. The rhombohedral unit cell vectors ($\mathbf{r}_1,\mathbf{r}_2,\mathbf{r}_3$) and the hexagonal unit cell vectors ($\mathbf{h}_1,\mathbf{h}_2,\mathbf{h}_3$) are related by the following transformation:

\begin{equation}
\left(
\begin{array}{ccc}
\mathbf{r}_1 \\
\mathbf{r}_2 \\
\mathbf{r}_3
\end{array}
\right)
=
\left(
\begin{array}{ccc}
-1 & 0 & 1 \\
 1 & 0 &-1 \\
 1 & 1 & 1
\end{array}
\right)
\left(
\begin{array}{ccc}
\mathbf{h}_1 \\
\mathbf{h}_2 \\
\mathbf{h}_3 
\end{array}
\right)
\end{equation}

In the modeled hexagonal cell, the material consists of three quintuple layers of atoms to resemble the LPE crystals. Each quintuple layer has two bismuth (Bi) and three tellurium (Te) atomic layers arranged in the sequence Te$_{(1)}$-Bi-Te$_{(2)}$-Bi-Te$_{(1)}$ \cite{luo_first-principles_2012}. The hexagonal cell is chosen in this work attending two main reasons: i) the transition from bulk to the exfoliated few layers systems is achieved by adding a vacuum layer along just one lattice vector, namely \textbf{h}$_3$, ii) supercell construction is needed to diminish the interactions between the periodic images of defects, this task being simpler with the hexagonal cell. Bonding between the quintuple layers is weak due to van der Waals forces between the Te$_{(1)}$ layers. Within the quintuple layer, the coupling is strong due to covalent or partially ionic bonds \cite{teweldebrhan_exfoliation_2010}. Here the subscripts (1) and (2) indicate two different chemical environments for the tellurium atoms. Each atom in the Te$_{(1)}$ layer is strongly bonded to three Bi atoms in the quintuple layer, while each atom in the Te$_{(2)}$ layer is octahedrally coordinated with six Bi atoms. In the quintuple layer, each Bi atom is bonded to three Te$_{(1)}$ atoms and to three Te$_{(2)}$ atoms, as can be seen in figure S1(a).\\

To consider the effects of point defects on the activation of magnetism in the material, we have used a 2$\times$2$\times$1 supercell constructed from the hexagonal unit cell (Figure S1). Each atomic layer in a quintuple layer has four atoms, so in each quintuple layer there are twelve tellurium atoms and eight bismuth atoms. Since there are three quintuple layers in the supercell, it contains a total of sixty atoms.\\

\subsubsection{DFT calculations.}
Density functional theory (DFT) calculations for Bi$_2$Te$_3$, both pristine and with defects, were performed using the Quantum-ESPRESSO simulation package \cite{giannozzi_quantum_2009} with norm-conserving pseudopotentials. The generalized gradient approximation (GGA) in the form of the Perdew-Burke-Ernzerhof (PBE) functional \cite{perdew_generalized_1996} was chosen to include electronic exchange-correlation interactions. Additionally, the energy cutoff for the plane waves was optimized up to a value of 120 Ry and for the charge density up to 720 Ry, as well as, a 5$\times$5$\times$3 Monkhorst-Pack mesh to sample the Brillouin zone. To correctly include the interactions between quintuple layers the van der Waals density functional (vdW-DF) \cite{thonhauser_van_2007} along with the Cooper exchange functional \cite{cooper_van_2010} have been used. This combination can faithfully reproduce the experimental distances between quintuple layers in unstressed Bi$_2$Te$_3$ \cite{luo_first-principles_2012}, due to the correct account of the exchange energy, which is missing in other studies. In the self-consistent calculations, the convergence threshold for the energy is set to $10^{-10}$ eV. In this study seven configurations for point defects, related to Bi-rich structures, were considered: 1) a Te$_{(1)}$ vacancy or deletion of an atom in this layer, denoted V$\mathrm{Te}_{(1)}$; 2) a Te$_{(2)}$ vacancy, denoted V$\mathrm{Te}_{(2)}$; 3) a
Te$_{(1)}$ anti-site defect or substitution of a Te atom in this layer by a Bi atom, denoted Bi-$\mathrm{Te}_{(1)}$; 4) a
Te$_{(2)}$ anti-site defect or substitution of a Te atom in this layer by a Bi atom, denoted Bi-$\mathrm{Te}_{(2)}$; 5) Two antisites Bi-$\mathrm{Te}_{(1)}$ and Bi-$\mathrm{Te}_{(2)}$; 6) vacancy V$\mathrm{Te}_{(1)}$ and antisite Bi-$\mathrm{Te}_{(2)}$; 7) vacancy V$\mathrm{Te}_{(2)}$ and antisite Bi-$\mathrm{Te}_{(1)}$. All internal atomic coordinates and lattice constants of all configurations were fully relaxed using polarized spin calculations, until the maximum component of the Hellmann-Feynman force acting on each ion is less than 0.003 eV/\AA.

\section{Results}

\begin{figure}
	\includegraphics[scale=0.175]{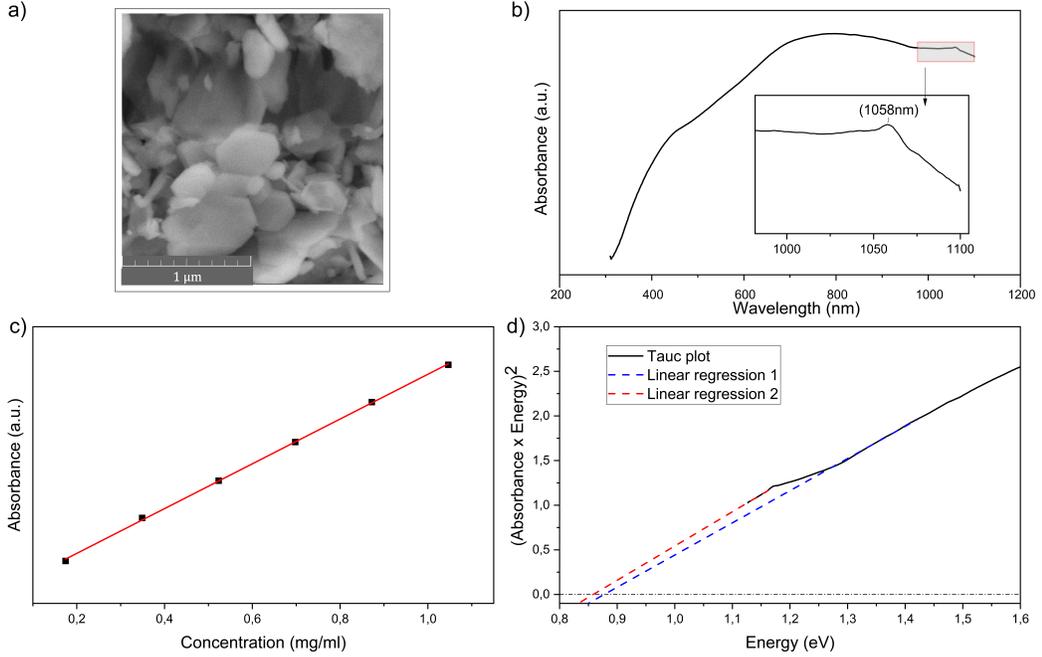}
	\centering
	\caption{a) SEM measurement; b) Exfoliated Bi$_2$Te$_3$ UV-Vis absorbance; (c) Exfoliated absorbance at $\lambda$ = 660 nm for solutions with different concentration resulting in an absorption coefficient of 182.06 ($\pm$2.75 \%) ml (mg$^{-1}$ m$^{-1}$); d) Exfoliated Bi$_2$Te$_3$ Tauc plot with intercept with the x-axis between 0.86 - 0.88 eV.}
	\label{fig:fig2}
\end{figure}

\subsection{Bi$_2$Te$_3$ characterization}

Bi$_2$Te$_3$ dispersions were prepared following the protocol in the methods section to achieve concentrations of 0.79 mg/ml. To confirm the success of the exfoliation process, Scanning Electron Microscopy (SEM) images were taken (Figure \ref{fig:fig2}a). Also, UV-Vis spectroscopy measurements were taken on the exfoliated Bi$_2$Te$_3$ in NMP/Water samples to characterize the optical response of the dispersion. Broad absorbance was observed from 300 to 1100 nm for the exfoliated crystals (Figure 1b) which is in good agreement with previously reported data \cite{zhang_few-layer_2020, zhao_actively_2016}. An absorbance vs concentration plot allowed the estimation of the optical absorbance coefficient at 660 nm resulting in 182.06 ($\pm$2.75 \%) ml (mg$^{-1}$ m$^{-1}$) (Figure 1c). This wavelength was chosen to compare it with other reported results for 2D materials. By employing the method proposed by Łukasz et al.\cite{harynski_facile_2022} and analyzing the measured UV-Vis data, the Tauc-plot exponent was determined as ($n = \frac{1}{2}$). This value corresponds to a semiconductor where the dominant transitions are direct. Consequently, we calculated Tauc-plots for a direct semiconductor using exfoliated $Bi_2Te_3$ dispersions based on absorption data (Figure 1d). For precise observation, the UV-Vis range for the bandgap calculation was selected from 1000 nm to 1100 nm (Figure 1b (inset)). This range is significant as literature indicates it corresponds to a region of Two-Photon Absorption (TPA) in Bi$_2$Te$_3$ \cite{qiao_two-photon_2019}. The resulting optical bandgap, ($E_g$), falls within the range of 0.86 eV to 0.88 eV, indicating that material exfoliation effectively tunes the bandgap. The measured value for this bandgap depends on the thickness distributions in the crystals, which vary in each sample (Figures S6 and S7).\\

\begin{figure}
	\includegraphics[scale=0.18]{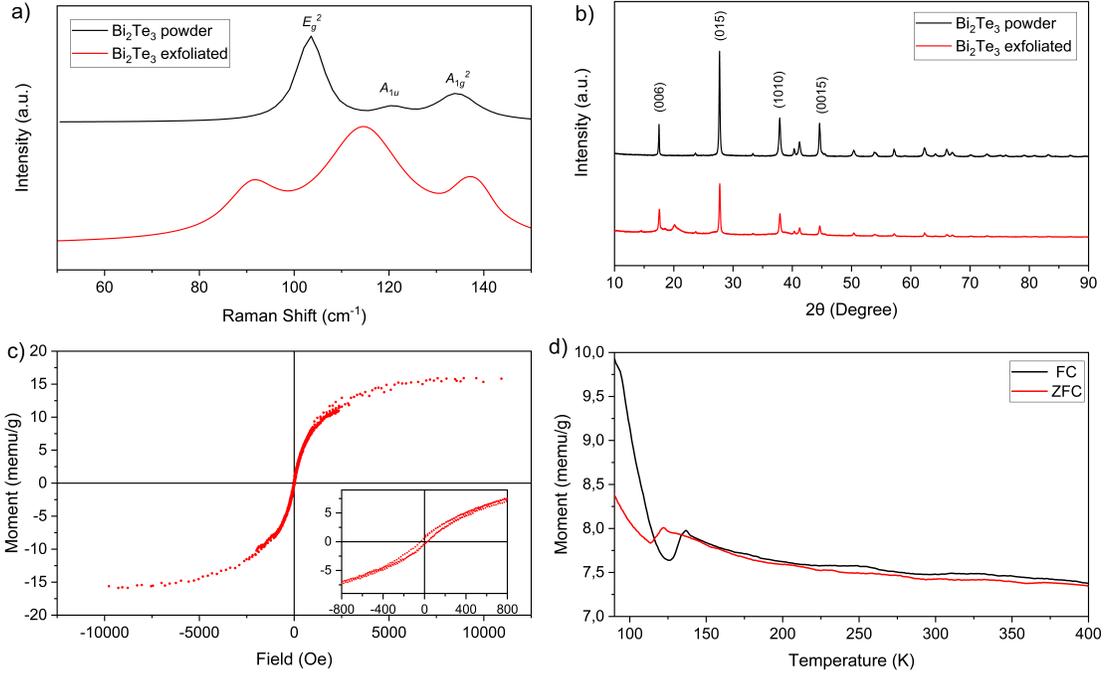}
	\centering
	\caption{a) Raman spectra and b) XRD spectra of the Bi$_2$Te$_3$ powder and exfoliated flakes; c) Hysteresis curve at room temperature and (d) Zero Field Cooled (ZFC) and Field Cooled (FC) magnetization curves of the Bi$_2$Te$_3$ exfoliated flakes.}
	\label{fig:fig_vivinitus}
\end{figure}

\begin{table}
\centering
\begin{tabular}{|l|l|l|l|l|l|l|}
\hline
\textbf{}           & \textbf{$E_g^{2}$} & \textbf{$A_{1u}$} & \textbf{$A_{1g}^{2}$} & \textbf{$I(A_{1g}^{2})/I(E_g^{2})$} & \textbf{$I(A_{1u})/I(E_g^{2})$} & \textbf{Ref.} \\ \hline
\textbf{Powder}     & 91.91        & 114.36       & 136.81        & 1.08                    & 2.45                   & This work     \\ \hline
\textbf{Exfoliated} & 103.4        & 120.6        & 134.3         & 0.31                    & 0.12                   & This work     \\ \hline
\textbf{Bulk}       & 101.7        & ---          & 134           & 0.75                    & ---                    & Shahil\cite{shahil_crystal_2010}      \\ \hline
\textbf{82nm}       & 101.9        & 116.9        & 132.7         & 0.83                    & 0.62                   & Shahil\cite{shahil_crystal_2010}      \\ \hline
\end{tabular}
\caption{Raman peaks and intensities for $Bi_2Te_3$ powder and exfoliated flakes.}
\label{tab:tableuvvis}
\end{table}

Raman spectra analysis for the Bi$_2$Te$_3$ powder and exfoliated flakes was performed using a 532nm laser (Figure \ref{fig:fig_vivinitus}a). Bi$_2$Te$_3$ has 15 lattice vibrational modes (12 optical and 3 acoustic). The 12 optical modes are described by the symmetries $2A_{2u} + 2E_u + 2A_{1g} + 2E_g$. For the flakes, three prominent Raman peaks are observed at 103.4 cm$^{-1}$, 120.6 cm$^{-1}$, and 134.3 cm$^{-1}$, corresponding to the $Eg^{2}$, $A_{1u}$, and $A_{1g}^{2}$ modes, respectively. Notably, the $A_{1u}$ peak becomes Raman-active only at low thicknesses due to the breaking of crystal symmetry \cite{shahil_crystal_2010}. These results are in agreement with those reported in the literature \cite{ren_large-scale_2012,shahil_crystal_2010}. The frequencies of the peaks observed for the powder and flakes, the relationship between intensities, and the values reported in previous reports are synthesized (Table \ref{tab:tableuvvis}). For the powdered material, a noticeable Raman shift is observed compared to the exfoliated sample. The $Eg^{2}$ and $A_{1u}$ peaks exhibit red shifts of 11.49 cm$^{-1}$ and 6.24 cm$^{-1}$, respectively, while the $A_{1g}^{2}$ peak shows a blue shift of 2.5 cm$^{-1}$. Additionally, the intensity ratio $I(A_{1g}^{2})/I(E_g^{2})$ is presented, as this ratio is expected to decrease as the material becomes thinner.\\

X-ray diffraction (XRD) analysis of the Bi$_2$Te$_3$ powder and exfoliated flakes was performed using an Empyrean Series 3 diffractometer from Malvern Panalytical. The spectra acquired is presented in (Figure \ref{fig:fig_vivinitus}b). The results confirm the phase purity of the material and are in agreement with previously reported data \cite{mansour2014structural}. The observed decrease in peak intensity for the exfoliated material compared to the bulk material can be attributed to the reduction in the number of layers and increased disorder, which diminishes the diffraction peaks. \\

\subsection{Bi$_2$Te$_3$ magnetic characterization}

Magnetic measurements of the Bi$_2$Te$_3$ exfoliated flakes were conducted using a Vibrating Sample Magnetometer (VSM) at room temperature (Figure \ref{fig:fig_vivinitus}c). The characteristic hysteresis curve confirms the presence of ferromagnetic behavior in the sample, with the material achieving magnetic saturation ($15.83$ memu/g) at relatively low magnetic fields. Additionally, the magnetic response of Bi$_2$Te$_3$ before exfoliation was measured, showing a diamagnetic signal (Figure S4), confirming that the ferromagnetism is a consequence of the exfoliation. The low-temperature magnetization behavior of exfoliated Bi$_2$Te$_3$ under zero-field-cooled (ZFC) and field-cooled (FC) conditions was also measured (Figure \ref{fig:fig_vivinitus}d). The fact that the FC curve lies above the ZFC curve at room temperature confirms the ferromagnetic nature of the sample. Additionally, this behavior is evidenced across the temperature range of 250K to 400K, indicating that the ferromagnetic phase is stable within this range. The stability across such a broad temperature interval suggests robustness in the underlying mechanisms driving the ferromagnetism, such as vacancy defects and antisites.\\

Based on the ZFC curve, a blocking temperature of 120 K is identified. Above this temperature, the spins are liberated, leading to the onset of ferromagnetic behavior, which stabilizes at temperatures above 250 K. This transition is likely driven by spin competition at low temperature (Figure S5), although a more detailed investigation of this phenomenon is beyond the scope of this work.

\subsection{DFT Results}

\begin{table}
\centering
\begin{tabular}{|l|c|c|c|c|c|c|}
\hline
\textbf{System} & \textbf{a(Å)} & \textbf{c(Å)} & \textbf{TM($\mu_B$/f.u.)} & \textbf{TM(memu/g)} & \textbf{Gap(eV)} & \textbf{d}$_{QL}$(\AA) \\ \hline
Bi$_2$Te$_3$\cite{wyckoff1964,chen_experimental_2009} & 4.384 & 30.487 & 0.00 & 0.00 & 0.170 & 2.612 \\ \hline
Bi$_2$Te$_3$ & 4.265 & 30.375 & 0.00 & 0.00 & 0.300 & 2.587 \\ \hline
BiTe$_{(1)}$ & 4.255 & 30.386 & 3.33$\times$ 10$^{-3}$ & 4.48 & metal & 2.526 \\ \hline
BiTe$_{(2)}$ & 4.256 & 30.340 & 5.83$\times$ 10$^{-3}$& 7.83 & metal & 2.531 \\ \hline
VTe$_{(1)}$ & 4.250 & 30.165 & 0.00 & 0.00 & 0.266  & 2.554 \\ \hline
VTe$_{(2)}$ & 4.251 & 30.176 & 0.00 & 0.00 & 0.178  & 2.478 \\ \hline
BiTe$_{(1)}$-BiTe$_{(2)}$ & 4.377 & 30.311 & 0.00 & 0.00& metal & 2.534 \\ \hline
BiTe$_{(1)}$-VTe$_{(2)}$ & 4.373 & 30.246 & 8.83$\times$10$^{-4}$& 1.12 & metal & 2.514 \\ \hline
BiTe$_{(2)}$-VTe$_{(1)}$ & 4.368 & 30.300 & 6.33$\times$10$^{-2}$ & 85.03 & metal & 2.579 \\ \hline
\end{tabular}
\caption{Optimized lattice parameters (a and c) in \AA, total magnetization (TM) in $\mu_B$/{f.u.} and memu/g, and bandgap in eV for pristine Bi$_2$Te$_3$ (which is compared to the experimental value) and for  configurations with defects. Last column displays the calculated van der Waals gap (distance between adjacent quintuple layers).}
\label{tab:lattice_parameters}
\end{table}

First principles calculations of defect induced magnetization in Bi$_2$Te$_3$ reported previously \cite{xiao2014unexpected} allowed to identify the Bi-Te$_{(2)}$ antisite as the defect that triggers magnetization. In the present work we have extended the aforementioned study to configurations with concurrent antisites and vacancies, making a systematic evaluation of the structure and magnetization for the resulting configurations. Spin-orbit coupling (SOC) was included for selected configurations aiming at the evaluation of possible effects of this interaction which can play an important role in topological insulators. Possible changes in both structural and magnetic properties of these materials upon exfoliation have been addressed by performing calculations for three isolated quintuple layers, (Figure S1).
Table \ref{tab:lattice_parameters} shows the fully optimized lattice constants, the total magnetic moment (TM) in $\mu_B$/f.u. and emu/g, and the band-gap between the valence band maximum and the conduction band minimum. The calculated lattice constants a=4.265 \AA~ and c=30.375 \AA~ are underestimated compared to the experimentally measured values a=4.384 \AA~ and c=30.487 \AA~ \cite{wyckoff1964}, however, they are close to the values a=4.360 \AA~ and c=30.174 \AA, obtained in reference \cite{luo_first-principles_2012} also using Cooper exchange; the difference may be due to the use of different convergence criteria and different pseudopotentials. Experimental band gap measurements for Bi$_2$Te$_3$ using different techniques have been reported, the values obtained are between 0.26 and 0.92 eV \cite{chen_experimental_2009,resendiz-hernandez_structural_2023}. From the present calculations it has been obtained a direct band gap at Gamma point of 0.304 eV. Although the value of this result is within the experimental range, it is known that the value of the band gap is underestimated when the PBE exchange-correlation potential is used. In addition, the underestimation of the experimental lattice constants, which has been presented in this study, may have contributed to the increase of the band gap value. The result reported here is slightly smaller than the value reported by Larson et al. (0.37 eV-direct at Gamma) \cite{larson2000} that was obtained using the same GGA exchange-correlation potential with a different method to expand the Kohn-Sham functions. In that case a linearized augmented plane wave (LAPW) full potential was used, and also the spin-orbit interaction was considered.
It can be seen that the tellurium-vacancy configurations V$\mathrm{Te}_{(1)}$ and V$\mathrm{Te}_{(2)}$ maintain the semiconducting character of the pristine compound. Thomas et al. have carried out optical measurements on Bi-doped Bi$_2$Te$_3$ at 10 K and found a gap of 0.150$\pm$0.02 eV at a weak absorption edge and other gap of 0.22$\pm$0.02 eV \cite{thomas1992}, which agree very well with the results in the present work for 
V$\mathrm{Te}_{(1)}$ (0.266 eV) and V$\mathrm{Te}_{(2)}$ (0.178 eV) at 0 K.
The presence of Bi antisite defects gives rise to a metallic character in the compound. Analysis of the density of states (DOS) plots and the resulting total magnetization indicate that Bi antisites are responsible for the activation of the magnetism in the Bi$_2$Te$_3$ compound. In particular, Bi$\mathrm{Te}_{(2)}$-V$\mathrm{Te}_{(1)}$ configuration shows a significant magnetization of 6.33$\times$10$^{-2}$ $\mu_B$/ f.u.=0.085 emu/g, which represents an increase of one order of magnitude if compared to the magnetization of the antisite Bi$\mathrm{Te}_{(2)}$. This means that the simultaneous presence of vacancy and antisite defects can provoke a synergic effect on the magnetization. The experimental measurement for the magnetization at 300 K is 0.016 emu/g, while the value measured at 5 K is 0.053 emu/g \cite{xiao2014unexpected}. The theoretical value here calculated at 0 K agrees well with the experimental value measured at 5 K, therefore present calculations confirm the experimental observation of magnetism in hexagonal Bi$_2$Te$_3$ activated by the point defect Bi$\mathrm{Te}_{(2)}$ together with the tellurium vacancy. Calculated electronic band structures for pristine Bi$_2$Te$_3$ and Bi$_2$Te$_3$ with Te antisites can be found in Figures S2 and S3 respectively. Regarding the stability of the magnetic phase in these materials, a comparison of the total energy of the ground state with and without spin polarization for Bi$\mathrm{Te}_{(2)}$ and Bi$\mathrm{Te}_{(2)}$-V$\mathrm{Te}_{(1)}$ configurations, show that in both cases the magnetic systems are more stable than the non-magnetic ones by 0.108 meV and 7.355 meV respectively.  \\

On the other hand, the distance between adjacent quintuple layers for all the systems considered has been calculated using the vdW-DF functional, with the Cooper exchange (vdW-DFC09x). The resulting distance of 2.587 \AA~ for Bi$_2$Te$_3$ is quite close to the value of 2.582 \AA~ reported in ref \cite{luo_first-principles_2012}, and is in good agreement with the experimental value of 2.612\AA~ \cite{nakajima1963}. It can be seen that the presence of defects causes the distance between adjacent quintuple layers to slightly decrease  compared to that of Bi$_2$Te$_3$. \\
\begin{figure}
    \centering
    \includegraphics[width=0.49\linewidth]{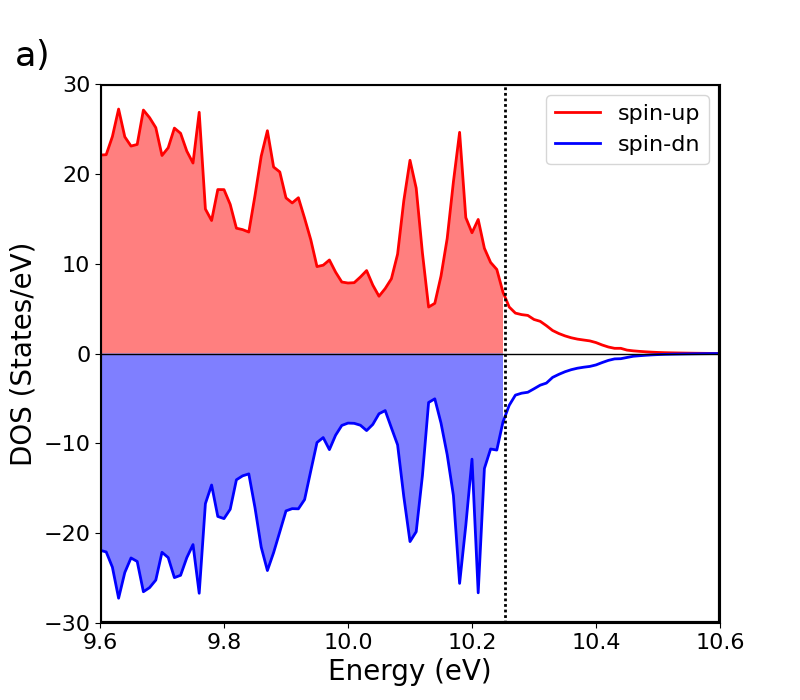}
    \includegraphics[width=0.49\linewidth]{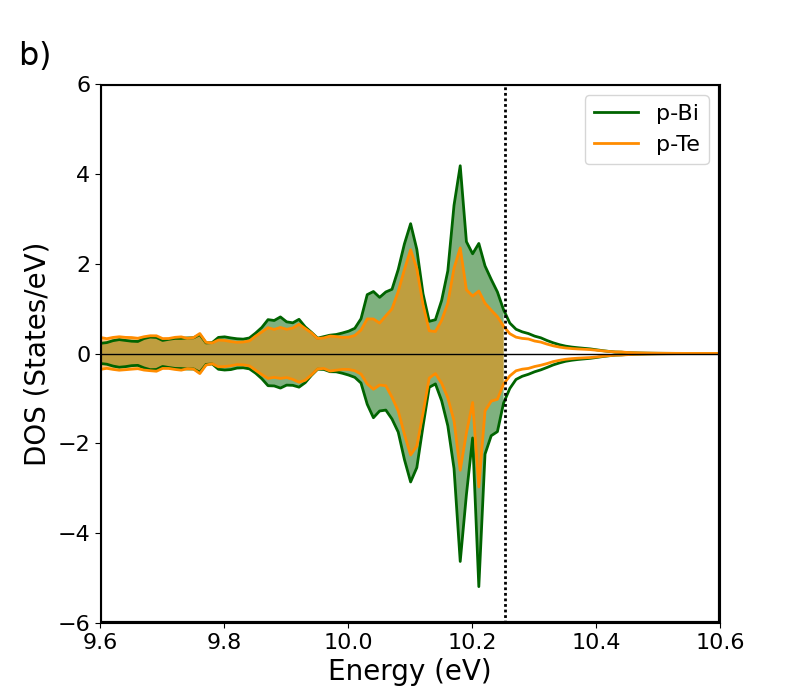}
    \includegraphics[width=0.49\linewidth]{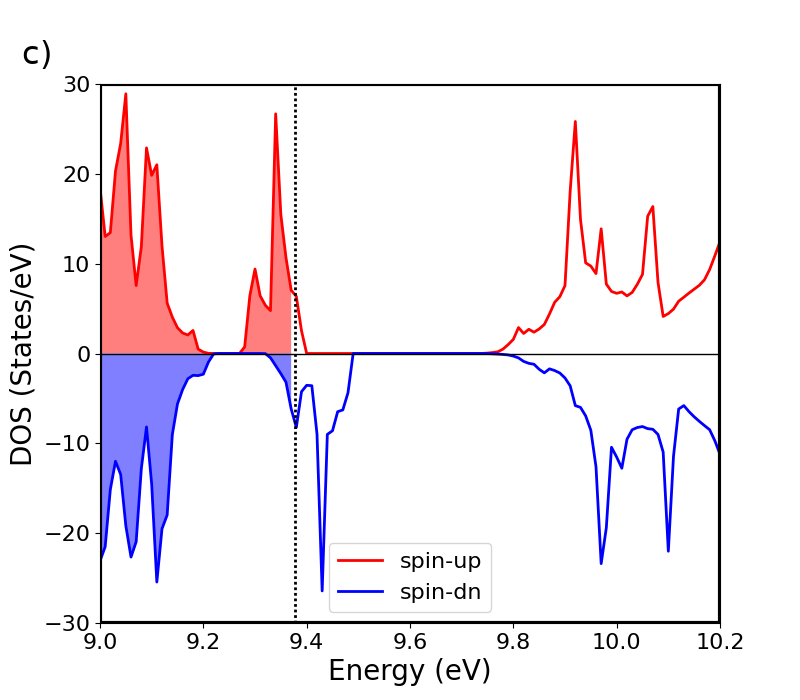}
    \includegraphics[width=0.49\linewidth]{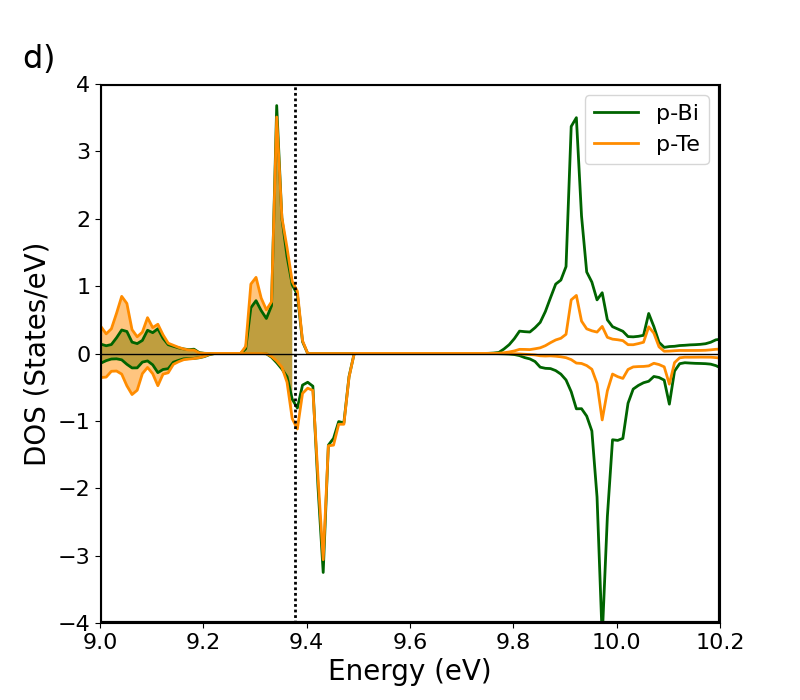}
    \caption{Calculated total spin-polarized density of states (DOS) for the two configurations which highlight the ferromagnetism in the defective Bi$_2$Te$_3$ compound. a) Total spin-polarized DOS for the Bi-Te$_{(2)}$ configuration, b) DOS projected onto the Bi and Te atoms that contribute most to the magnetization in the Bi-Te$_{(2)}$  configuration, c) total spin-polarized DOS for the Bi-Te$_{(2)}$-VTe$_{(1)}$ configuration, d) DOS projected onto the Bi and Te atoms that contribute most to the magnetization in the Bi-Te$_{(2)}$-VTe$_{(1)}$ configuration. The vertical dashed line indicates the position of the Fermi level.}
    \label{pdos}
\end{figure}

In order to obtain information on the origin of ferromagnetism in the aforementioned systems, calculations of the total density of states (DOS) and projected density of states were performed for the two configurations with the highest magnetization. In Figure \ref{pdos}a, the DOS for the Bi-Te$_{(1)}$ configuration is shown and it can be seen that the up-spin and down-spin curves are asymmetric in the valence band near the Fermi level, indicating the presence of a ferromagnetic state in the system with this particular defect. The density of states projected on the anti-sitiated Bi atom and on the Te atom which contribute most to the magnetization are presented in Figure \ref{pdos}b. The local magnetic moment in each atom is due to the p-orbitals. The ferromagnetism in the anti-sitiated Bi-Te$_{(2)}$ compound originates from the hybridization of the p-orbitals belonging to the two Bi and Te atoms which contribute most to the magnetization, and which are located in the same quintuple layer. In the Bi-Te$_{(2)}$ antisite configuration plus a tellurium vacancy VTe$_{(1)}$ (see Figure \ref{pdos}c), an asymmetric density of states (DOS) is observed for the spin-up and spin-down curves, mainly around the Fermi level. A higher asymmetry if compared to the Bi antisite configuration without tellurium vacancy can be seen, which is in agreement with the finding of a higher magnetization and a more ferromagnetic state in the defective Bi-Te$_{(2)}$-VTe$_{(1)}$ system. Figure \ref{pdos}d shows the DOS projected on the Bi and Te atom belonging to the same quintuple layer which contributes the most to the magnetization. It can be seen that the p orbitals are responsible for the local magnetization in each of these two atoms. The ferromagnetism in the Bi-Te$_{(2)}$-VTe$_{(1)}$ configuration originates from a strong hybridisation of the p orbitals belonging to these two atoms of Bi and Te.\\ 

Simulations of three isolated quintuple layers were also performed. In this case a 30 \AA~ vacuum in the z-direction was included, as shown in figure S1(b). For the defect-free slab a band gap of 0.6 eV is found, which is in good agreement with the experimental measurement (0.8 eV) obtained in the present work. For this case the system remains non-magnetic. As before, configurations with the two highest bulk magnetization were chosen, BiTe$_{(2)}$ and BiTe$_{(2)}$-VTe$_{(1)}$, which were modeled using the three-quintuple slab. The slab with the  BiTe$_{(2)}$ configuration turns out to be ferromagnetic with a magnetization of 0.44 $\mu_B$/cell = 0.049 emu/g, which is higher than that of this configuration in the bulk. The BiTe$_{(2)}$-VTe$_{(1)}$ configuration in the three isolated quintuple layers turns out to be ferromagnetic with a magnetization of 0.6 $\mu_B$/cell = 0.067 emu/g, which is slightly lower than that obtained in bulk. The resulting magnetization values are in good agreement with the experimental measurements reported in \cite{xiao2014unexpected} and with those measured in the present work. For each case, the slabs were built from the relaxed structures that were found in the corresponding bulk systems.\\ 

Possible effects of the Hubbard potential in these systems have been ruled out by DFT+U calculations of the magnetization and electronic band gap using estimated Hubbard parameters in the 0.1 eV to 1.0 eV range. This parameter was included for the p orbital of the Bi atom in the antisite BiTe$_{(2)}$ given its large contribution to the magnetization. The results confirm a negligible effect on both the magnetization and the band gap, as the metallic character of the material is preserved in the full range of studied Hubbard parameters. These results can be expected from the electronic configurations of Te and Bi, which lack of d or f localized electrons in their valence states.

\section{Conclusions}

A detailed characterization of exfoliated  Bi$_2$Te$_3$ dispersions produced via solvothermal intercalation was conducted, demonstrating for the first time that the material exhibits room-temperature ferromagnetism following liquid phase exfoliation. Our complementary experimental and computational research suggests that this magnetic behavior arises due to the presence of vacancies and antisites, specifically in the BiTe$_{(2)}$ and BiTe$_{(2)}$-VTe$_{(1)}$ configurations, within both the bulk material and the exfoliated nanostructures. Computational modeling reveals ferromagnetism in the metallic configuration, while experimental results demonstrate both ferromagnetic properties and semiconducting optical behavior, suggesting a polydisperse mixture. Additionally we may conclude that as the material becomes thinner (with a narrow distribution of number of layers), a broadening of the optical bandgap occurs, as measured by UV-Vis spectroscopy and obtained by computational modelling. These findings indicate that liquid-phase exfoliation not only alters the electronic properties of Bi$_2$Te$_3$, but also offers new possibilities for tuning its magnetic characteristics, making it a promising candidate for future applications in spintronics and optoelectronics.

\section{Acknowledgements}

C. Espejo and W. López acknowledge computational resources from
Granado-HPC, Universidad del Norte. Y. Hernandez acknowledge funding by the Faculty of Sciences at Universidad de los Andes within project INV-2023-162-2843. 

\section*{References}

\bibliographystyle{unsrt}
\bibliography{lib.bib}

\end{document}